\documentclass[a4paper,11pt]{article}
\usepackage{pos}
\usepackage{subcaption}
\usepackage{CJKutf8}
\usepackage{enumitem}
\usepackage{comment}

\usepackage[scaled=1.15]{urwchancal}
\DeclareFontFamily{OT1}{pzc}{}
\DeclareFontShape{OT1}{pzc}{m}{it}%
{<-> s * [1.15] pzcmi7t}{}
\DeclareMathAlphabet{\mathpzc}{OT1}{pzc}{m}{it}

\begin{document}

\title{Pion and Kaon Fragmentation Functions from Continuum Schwinger Function Methods}


\author*[a,b]{Hui-Yu Xing%
    $\,^{\href{https://orcid.org/0000-0002-0719-7526}{\textcolor[rgb]{0.00,1.00,0.00}{\sf ID}},}$}

\affiliation[a]{School of Physics, \href{https://ror.org/01rxvg760}{Nanjing University},
Nanjing, Jiangsu 210093, China}

\affiliation[b]{Institute for Nonperturbative Physics, \href{https://ror.org/01rxvg760}{Nanjing University}, Nanjing, Jiangsu 210093, China}

\emailAdd{hyxing@nju.edu.cn}

\abstract{Using the Drell–Levy–Yan relation, the pion and kaon elementary fragmentation functions (EFFs) are obtained from their hadron-scale parton distribution functions (DFs).  These EFFs serve as driving terms in the hadron cascade equations, whose
solution yields the complete array of hadron-scale fragmentation functions (FFs) for pion and kaon production in high
energy reactions.  Evolved to experimental scales, the continuum Schwinger function methods (CSMs) predictions satisfy QCD endpoint behavior: nonsinglet FFs vanish at $z=0$, singlet FFs diverge faster than $1/z$. Jet multiplicity predictions reveal SU(3) symmetry breaking in the charged/neutral kaon ratio, decreasing with energy, and show the pion/kaon ratio in $e^+e^-$ collisions asymptotes to a mass-independent value.}

\FullConference{The 26th International Symposium on Spin Physics (SPIN25) -- A Century of Spin\\
2025 September 22-26\\
Qingdao Haitian Hotel, Qingdao, Shandong Province, China\\
and\\
The 2025 International Conference on the Structure of Baryons (Baryons 2025)\\
2025 November 10-14\\
the International Convention Center (ICC), Jeju, Korea
}


\maketitle

\section{Introduction}

Jets of energetic hadrons are often produced in high energy interactions. Within the framework of Quantum Chromodynamics (QCD), they are understood to originate from quark and gluon partons, which, after being produced in the initial collision, escape the interaction region and, under the influence of confinement dynamics, fragment into a shower of colorless hadrons \cite{Field:1976ve, Field:1977fa, Altarelli:1981ax, Ellis:1991qj, Metz:2016swz, Chen:2023kqw}. This nonperturbative hadronization process is quantified by fragmentation functions (FFs), denoted as $D_q^h(z; \zeta)$. FFs describe the number density for a parton $q$, at a resolving scale $\zeta$, to produce hadrons $h$ carrying a light-front momentum fraction $z$.

Precise knowledge of FFs is increasingly critical. They are indispensable ingredients in the extraction of Parton Distribution Functions (DFs) and play a central role in the physics programs of current and future facilities \cite{Aguilar:2019teb, BESIII:2020nme, Anderle:2021wcy, Arrington:2021biu, Schnell:2022nlf, Quintans:2022utc}. Traditionally, FFs are determined via phenomenological fits \cite{Hirai:2007cx, deFlorian:2014xna, Bertone:2017tyb, Soleymaninia:2020bsq, Moffat:2021dji, AbdulKhalek:2022laj, Gao:2024dbv} to experimental data assuming collinear factorization. However, these extractions often suffer from large uncertainties, especially in the kinematic endpoint regions ($z \simeq 0,1$) where data is sparse. Furthermore, FFs are nonperturbative quantities, and few realistic calculations have been available. Owing to their timelike nature, lattice-regularized QCD is ill-suited to FF computation. An alternative approach is provided by continuum Schwinger function methods (CSMs)~\cite{Roberts:2021nhw, Binosi:2022djx, Ding:2022ows, Ferreira:2023fva, Raya:2024ejx}, with a recent application to pion FFs given in Ref.~\cite{Xing:2023pms}.

Within CSMs, FFs can be computed by exploiting the Drell–Levy–Yan (DLY) relation \cite{Drell:1969jm, Drell:1969wd, Drell:1969wb, Gribov:1972ri, Gribov:1972rt}. The DLY relation suggests a crossing symmetry that connects the DFs of a hadron, ${\mathpzc q}^h(x)$, to the elementary fragmentation functions (EFFs), $d_{\mathpzc q}^h(z)$, at the hadron scale $\zeta_{\cal H}$:
\begin{equation}
    d_{\mathpzc q}^h(z;\zeta_{\cal H}) = z \, {\mathpzc q}^h\left(\frac{1}{z} ;\zeta_{\cal H}\right)\,.
    \label{DLYR}
\end{equation}

Eq.\,(\ref{DLYR}) implies that all manifestations of emergent hadron mass (EHM) \cite{Roberts:2021nhw, Binosi:2022djx, Ding:2022ows, Ferreira:2023fva, Raya:2024ejx, Salme:2022eoy, Krein:2023azg} present in DFs are directly mapped into the source function driving the fragmentation. Consequently, perhaps, the seeds of confinement, as expressed in hadronization, can already be found in the wave functions of the hadrons involved.

\begin{figure*}[t]
    \centering
    \setlength{\tabcolsep}{12pt}
\begin{tabular}{cc}
{\sf A} & {\sf B}  \\
\includegraphics[clip, width=0.4\columnwidth]{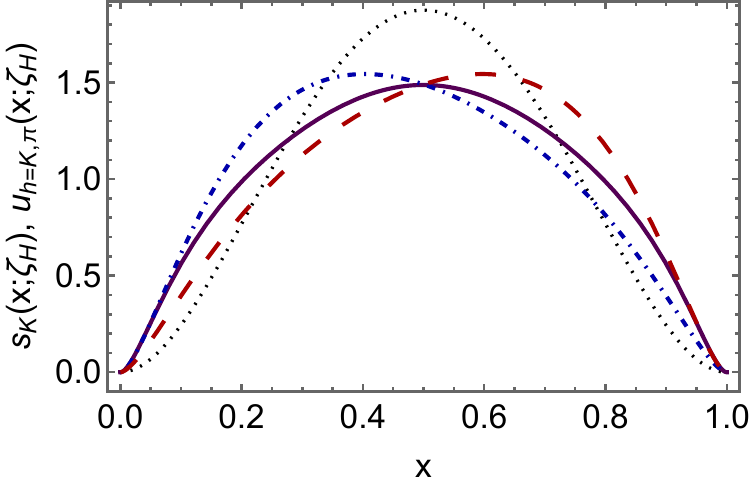} &
\includegraphics[clip, width=0.4\columnwidth]{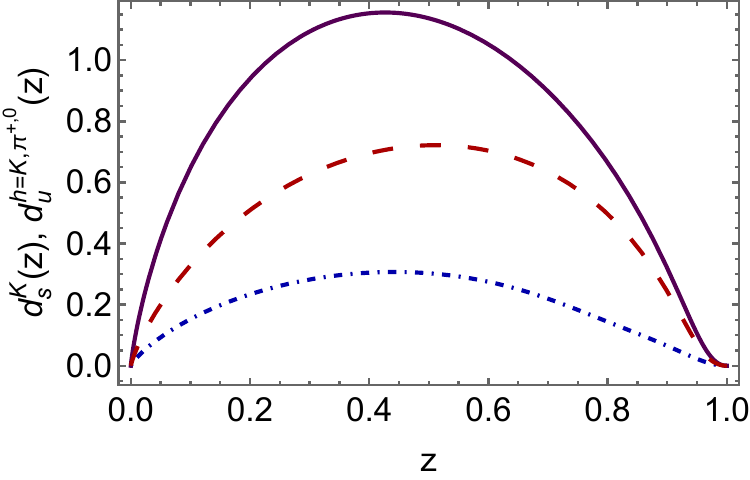} 
\end{tabular}
    \vspace*{0ex}
\caption{\label{CSMpiKDFs}
{\sf Panel A}.
Dressed valence quark parton distribution functions evaluated using CSMs in Ref.\,\cite{Cui:2020tdf}:
${\mathpzc s}_{K^-}(x ; \zeta_H)$ -- long-dashed red curve;
${\mathpzc u}_{K^+}(x ; \zeta_H)$ -- dot-dashed blue;
${\mathpzc u}_{\pi^+}(x ; \zeta_H)$ -- solid purple;
scale-free DF -- dotted black.
{\sf Panel B}.
Realistic elementary fragmentation functions, obtained from the $\pi, K$ curves in {\sf Panel A} using Eq.\,\eqref{DLYR}.
$d_{\mathpzc s}^{K^-}(x ; \zeta_H)$ -- long-dashed red curve;
$d_{\mathpzc u}^{K^+}(x ; \zeta_H)$ -- dot-dashed blue;
$d_{\mathpzc u}^{\pi^++\pi^0}(x ; \zeta_H)$ -- solid purple.
}
\end{figure*}

\section{Elementary fragmentation functions}

The pion and kaon dressed valence quark DFs at hadron scale were given in Ref.\,\cite{Cui:2020tdf}.  Drawn in Fig.\,\ref{CSMpiKDFs}\,A, they may be represented by the following functions:
{\allowdisplaybreaks
\begin{subequations}
\label{piKDFs}
\begin{align}
{\mathpzc u}_\pi(x;\zeta_{\cal H}) & =
{\mathpzc n}_\pi \ln \left[1 + \tfrac{1}{\rho_\pi^2} x^2(1-x)^2  (1+ \tfrac{1}{2} \gamma_\pi^2 [([1-x]^2)^{\beta_\pi} +(x^2)^{\beta_\pi}])
\rule{0em}{2.2ex}\right]\,,
\end{align}
${\mathpzc n}_\pi = 0.858$,
$\rho_\pi = 0.116$,
$\gamma_\pi = 1.967$,
$\beta_\pi = 5.938$;
and
\begin{align}
{\mathpzc u}_K(x;\zeta_{\cal H}) & =
{\mathpzc n}_K \ln \left[ 1+\tfrac{1}{\rho_K^2} x^2 (1-x)^2  (1 + \gamma_K^2 (x^2)^{\alpha_K} ([1-x]^2)^{\beta_K}) \rule{0em}{2.3ex}\right]\,,
\end{align}
\end{subequations}
${\mathpzc n}_K = 0.444$,
$\rho_K = 0.0746$,
$\gamma_K = 6.276$,
$\alpha_K = 0.710$,
$\beta_K = 1.650$.
One has ${\mathpzc s}_{K^-}(x;\zeta_{\cal H}) = {\mathpzc u}_{K}(1-x;\zeta_{\cal H})$.

Owing to EHM, both the pion and kaon DFs are significantly dilated with respect to the scale-free DF, i.e., ${\mathpzc q}_{\rm sf} = 30 x^2 (1-x)^2$. In addition, the skewing for kaon DFs is the expression of the interference between EHM and  Higgs boson couplings into QCD. 

Using Eq.\,\eqref{DLYR} and the DFs in Fig.\,\ref{CSMpiKDFs}\,A, one obtains the EFFs drawn in Fig.\,\ref{CSMpiKDFs}\,B.
Since a $u$ quark can directly produce $\pi^{+}, \pi^{0}$ and $K^+$, then, the associated EFFs are normalized as 
\begin{equation}
\int_0^1 dz \, \left[ \tfrac{3}{2} d_{\mathpzc u}^{\pi^+}(z;\zeta_{\cal H}) + d_{\mathpzc u}^{K^+}(z;\zeta_{\cal H})\right] = 1.
\end{equation}

\section{Realistic fragmentation functions}

We follow Ref.\,\cite{Field:1977fa} in building complete FFs from EFFs.  Namely, with the EFF describing the first fragmentation event for parton $p$ generating hadron $h$ with momentum fraction $z$, then the complete FF, $D_{\mathpzc p}^h(z)$, is obtained via a recursion relation that resums the exhaustive series of such events:
\begin{equation}
D_{\mathpzc q}^h(z) =
d_{\mathpzc q}^h(z) + \sum_{{\mathpzc q}^\prime = {\mathpzc u}, {\mathpzc d}, {\mathpzc s}}\int_z^1 (dy/y)
d_{\mathpzc q}^{{\mathpzc q}^\prime}(z/y) D_{{\mathpzc q}^\prime}^h(y)\,,
\label{JetEq}
\end{equation}
where $h=\pi^\pm, \pi^0, K^\pm, K^0, \bar K^0$.
In all these equations, as explained in Ref.\,\cite{Xing:2023pms}, the resolving scale $\zeta=\zeta_{\cal H}$. The explicit form of Eq.\,(\ref{JetEq}) is given by Eq.\,(18) in Ref.\,\cite{Xing:2025eip}.

It is worth highlighting some features of the solutions to Eq.\,\eqref{JetEq}: (\textit{i}) $D_{\mathpzc p}^h(z) \stackrel{z\simeq 1}{\approx} d_{\mathpzc p}^h(z)$, because if the parton gives all its momentum to $h$, then there is none left to contribute to a cascade. (\textit{ii}) For a given parton species $p$, $\sum_h \int_0^1 dz\,z\, D_p^h(z) = 1$, where the sum runs over all hadrons contained in the shower.  This identity means that the hadron jet generated by the parton $p$ contains all the momentum of that initial state, neither more nor less. (\textit{iii}) $D_{\mathpzc p}^h(z) \stackrel{z \simeq 0}{=} \frac{1}{z}$ \cite{Field:1976ve},
because it costs nothing to produce hadrons with zero fraction of the initial parton momentum.  In practice, the impact of this infrared divergence is tamed by hadron masses.

\begin{figure*}[t]
\hspace*{-1ex}
\begin{tabular}{ccc}
{\sf A} & {\sf B} & {\sf C} \\
\includegraphics[clip, width=0.32\columnwidth]{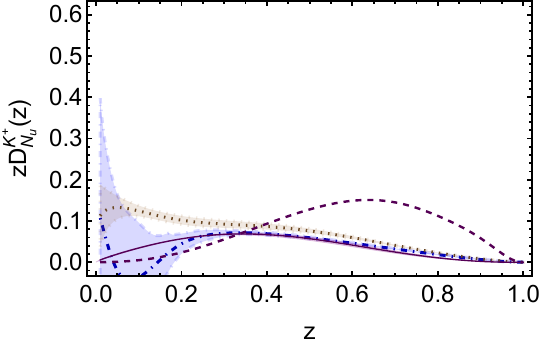} &
\includegraphics[clip, width=0.32\columnwidth]{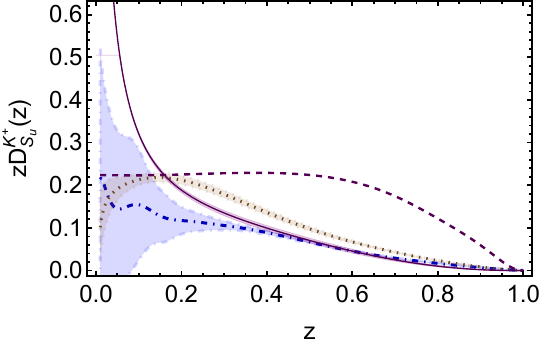} &
\includegraphics[clip, width=0.32\columnwidth]{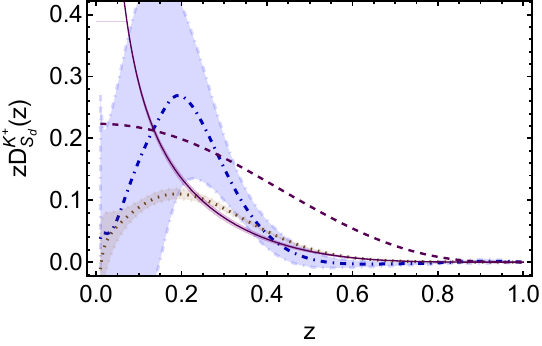}
\end{tabular}
\begin{tabular}{ccc}
{\sf D} & {\sf E} & {\sf F} \\
\includegraphics[clip, width=0.32\columnwidth]{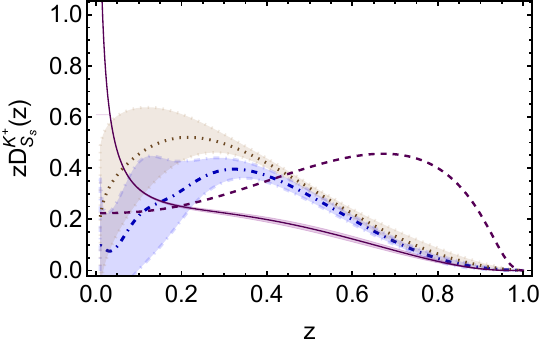} &
\includegraphics[clip, width=0.32\columnwidth]{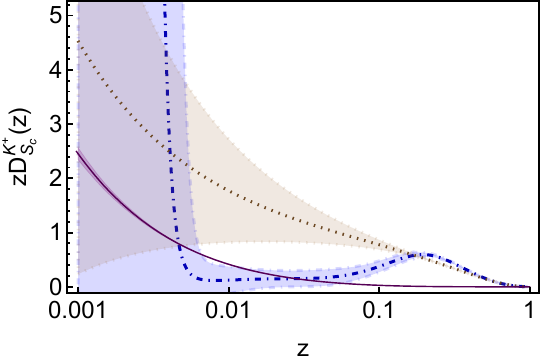} &
\includegraphics[clip, width=0.32\columnwidth]{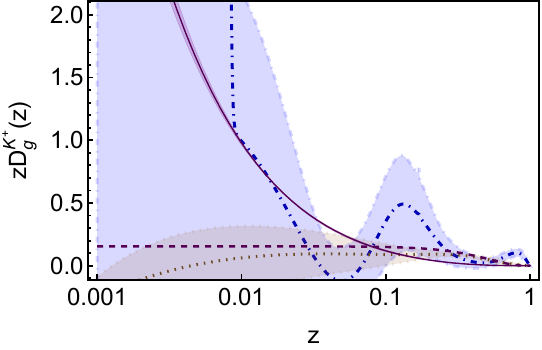}
\end{tabular}
\caption{\label{FCSMJetK}
CSM results for kaon fragmentation functions, defined in Eqs.\,\eqref{SK}\,--\,\eqref{NKs}.
Solutions of cascade equations, Eq.\,\eqref{JetEq} -- dashed purple curves.
All-orders evolution of those curves to $\zeta=\zeta_2 := 2\,$GeV -- solid purple curves, with uncertainty bands obtained as described in Sect.\,(5.2) in Ref.\,\cite{Xing:2025eip}.
Comparison curves are inferences from:
high-energy lepton-lepton, lepton-hadron and hadron-hadron scattering data \cite[JAM]{Moffat:2021dji} -- dotted brown curves, within like colored bands;
and electron-positron annihilation and lepton-nucleon semi-inclusive deep-inelastic scattering data \cite[MAPFF]{AbdulKhalek:2022laj} -- dot-dashed blue curves within like-colored bands.}
\end{figure*}

One works with the following singlet ($S$) and nonsinglet ($N$) combinations. Here, use the kaon as an example:
{\allowdisplaybreaks
\begin{subequations}
\label{EvolutionCombinations}
\begin{align}
D_{S_{\mathpzc q}}^{K^+}(z) & =  D_{\mathpzc q}^{K^+}(z) + D_{\bar{\mathpzc q}}^{K^+}(z)\,, \label{SK}\\
D_{N_{{\mathpzc q}\neq s}}^{K^+}(z) & =   D_{\mathpzc q}^{K^+}(z) - D_{\bar{\mathpzc q}}^{K^+}(z)\,,
\label{NKq} \\
D_{N_{{\mathpzc s}}}^{K^+}(z) & =   D_{\bar {\mathpzc s}}^{K^+}(z) - D_{{\mathpzc s}}^{K^+}(z)\,,
\label{NKs}
\end{align}
\end{subequations}
}

Our predictions for the kaon FFs are drawn in Fig.\,\ref{FCSMJetK}. For comparison, the inferences from data reported in Refs.\,\cite{Moffat:2021dji, AbdulKhalek:2022laj} are also shown.
One can note that the fits are mutually incompatible on $z\lesssim 0.5$.
Compared with our predictions, the situation is equally poor; namely, there is little agreement. The analogous comparison for the pion FFs leads to similar conclusions and is therefore not shown. Here, we consider the kaon FFs:
\begin{description}
[leftmargin=!,labelsep=0.5em,itemsep=4pt,topsep=6pt,parsep=0pt,partopsep=0pt]
  \item[Figs.\,\ref{FCSMJetK}\,A, B.]  $u \to K$  (favoured), nonsinglet and singlet.  There is agreement only on $z\gtrsim 0.4$, \emph{i.e}., on the valence quark domain.
      Here, the JAM and MAPFF results for $z D_N$ are finite and nonzero on $z\simeq 0$, which is unexpected. This is the domain of glue and sea dominance; so given Eq.\,\eqref{NKq}, $z D_N$ should vanish. 
      Moreover, $z D_S$ is also nonzero and finite, in contradiction with the CSM prediction.

\item[Figs.\,\ref{FCSMJetK}\,C,] $d \to K$.  One might claim qualitative agree\-ment on the far valence domain, but only because this FF is small. Both JAM and MAPFF produce nonzero finite values on $z\simeq 0$, where, on physics grounds, such outcomes are not expected. 

  \item[Figs.\,\ref{FCSMJetK}\,D] $s\to K$ (favoured), singlet.  Agreement is seen on $z\gtrsim 0.7$; but nothing beyond that. 
  Furthermore and once more unexpectedly, JAM and MAPFF fits produce non\-zero finite values on $z\simeq 0$.  Naturally, CSM predictions diverge on this glue and sea dominated domain.
  \item[Figs.\,\ref{FCSMJetK}\,E, F.] $c, g \to K$.  Quantitatively, there is no agreement on these FFs, which are very poorly constrained by data. $zD_{g}^{K^+}(z)$ from JAM is even negative on $z\simeq 0$, which is unphysical. On the other hand, the divergent behavior of $zD_{S_c}^{K^+}(z)$ from JAM, MAPFF and $zD_{g}^{K^+}(z)$ from MAPFF on $z\simeq 0$ is consistent with the CSM prediction.
\end{description}

It is worth highlighting that the non-monotonic (oscillatory) behavior of the MAPFF fits on $z\lesssim 0.5$ is entirely incompatible with our predictions.  Indeed, quite generally, the MAPFF results strongly suggest that FFs are practically unconstrained on $z\lesssim 0.2$.  The observations and remarks collected here indicate that, today, phenomenology does not deliver objective FF results: the results obtained are practitioner specific.

\section{Hadron jet multiplicities}

In the context of realizable experiments, consider $e^+ e^- \to h X$. An associated multiplicity structure function is normally defined as follows \cite[Sec.\,3.1.1]{Metz:2016swz}:
\begin{equation}
    F^h(z;\zeta)
    = \frac{1}{\sigma_{\rm tot}}\frac{d\sigma^{e^+e^- \to h X}}{dz}\,,
\end{equation}
with $\sigma_{\rm tot}$ being the total cross-section, so that $F^h(z;\zeta)$ is the number of $h$ hadrons produced in each event.
%
At leading order in perturbative QCD, $d\sigma^{e^+e^- \to h X}/dz
= \sum_{\mathpzc q}e_{\mathpzc q}^2 D_{\mathpzc q}^h(z;\zeta)$
and $\sigma_{\rm tot} = \sum_{q}e_q^2 =: \sigma_{\rm tot}^{\mbox{\rm\tiny  LO}}$.  
In terms of the multiplicity structure function, the total multiplicity is:
\begin{equation}
\label{FFmultzmin}
M^h(\zeta) = \int_{z_{\rm min}}^1 dz\, F^h(z;\zeta)=\int_{z_{\rm min}}^1 dz \frac{1}{\sigma_{\rm tot}}\frac{d\sigma^{e^+e^- \to h X}}{dz}= \int_{z_{\rm min}}^1 dz \frac{\sum_{\mathpzc q}e_{\mathpzc q}^2 D_{\mathpzc q}^h(z;\zeta)}{\sum_{q}e_q^2}\,.
\end{equation}

\begin{figure}[t]
\centerline{%
\includegraphics[clip, width=0.4\linewidth]{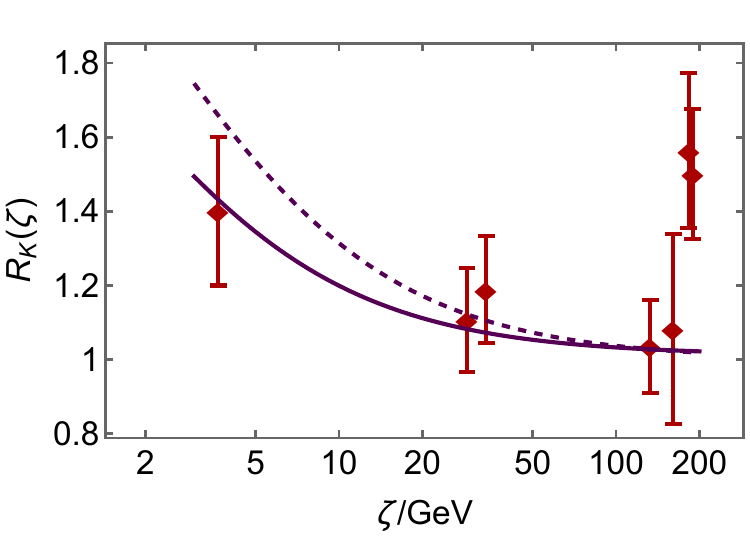}}
\caption{\label{FigRK}
$\zeta$-dependence of the charged/neutral kaon multiplicity ratio in Eq.\,\eqref{RKzeta0}. CSM prediction -- solid purple curve; SCI prediction in Ref.\,\cite{Xing:2025eip} -- dashed purple curve.
Data are empirical estimates from Refs.\,\cite{BESIII:2025mbc, TPCTwoGamma:1983lrv, TPCTwoGamma:1984eoj, TASSO:1988jma, DELPHI:2000ahn}.
}
\end{figure}

Conversion between experimental kinematics and $z$ is typically achieved by defining $z = 2  E_h/\sqrt{Q^2}$, where $Q^2=\zeta^2$ is the momentum transfer provided by the $e^+ e^-$ collision. It is clear that the minimum available value of the fragmentation momentum fraction is $z_{\rm min} = 2m_h/\zeta$, where $m_h$ is the mass of the produced hadron. $m_h$ places a natural lower bound on the integral in Eq.\,\eqref{FFmultzmin}.

To proceed, the charged/neutral kaon multiplicity ratio can be calculated and is shown in Fig.\,\ref{FigRK}:
\begin{equation}
\label{RKzeta0}
R_K(\zeta) = \frac{M^{K^+} (\zeta)+ M^{K^-} (\zeta)}{M^{K^0} (\zeta)+ M^{\bar K^0} (\zeta)}=\frac{\int_{z_{\rm min}}^1 dz\, \big[4 D_{S_{\mathpzc u}}^{K^+}(z;\zeta) 
+D_{S_{\mathpzc d}}^{K^+}(z;\zeta)+D_{S_{\mathpzc s}}^{K^+}(z;\zeta) + 4 D_{S_{\mathpzc c}}^{K^+}(z;\zeta)\big]}{\int_{z_{\rm min}}^1 dz\, \big[4 D_{S_{\mathpzc d}}^{K^+}(z;\zeta) 
+D_{S_{\mathpzc u}}^{K^+}(z;\zeta)+D_{S_{\mathpzc s}}^{K^+}(z;\zeta) + 4 D_{S_{\mathpzc c}}^{K^+}(z;\zeta)\big]}.
\end{equation}
Since $D_{S_{\mathpzc u}}^{K^+}(x;\zeta) \neq D_{S_{\mathpzc d}}^{K^+}(x;\zeta) = D_{S_{\mathpzc u}}^{K^0}(x;\zeta)$,
see Fig.\,\ref{FCSMJetK}\,B \emph{cf}.\ Fig.\,\ref{FCSMJetK}\,C, then $R_K(\zeta) \neq 1$.

In reality, $\sigma_{\rm tot}$ can only be calculated nonperturbatively and the measured value depends on many things, including experimental setup, detector acceptance, etc. Thus, in subsequent comparisons with experimental results for the $z$-dependent multiplicity distribution, we write 
\begin{equation}
F^h(z;\zeta) = \frac{1}{{\cal N}}
\sum_{\mathpzc q}e_{\mathpzc q}^2 D_{\mathpzc q}^h(z;\zeta)\,,
\label{FhNorm}
\end{equation}
with ${\cal N}$ chosen so as to ensure a fair match between our prediction and experiment for the integrated multiplicity over the central domain $z\in[0.1,0.7]$.  We choose this domain because, in our view, the experimental uncertainty in extant experimental results on the complement of this domain, \emph{i.e}., the deep sea and far valence domains, are underestimated. Here, $1/{\cal N} = 4.0(1.0) /\sigma_{\rm tot}^{\mbox{\rm\tiny  LO}}$, and with Eq.\,\eqref{FhNorm}, one obtains the pion and kaon results drawn in Fig.\,\ref{zdepmult}.

\begin{figure*}[t]
\hspace*{-1ex}
\setlength{\tabcolsep}{12pt}
\begin{tabular}{cc}
{\sf A} & {\sf B}  \\
\includegraphics[clip, width=0.4\columnwidth]{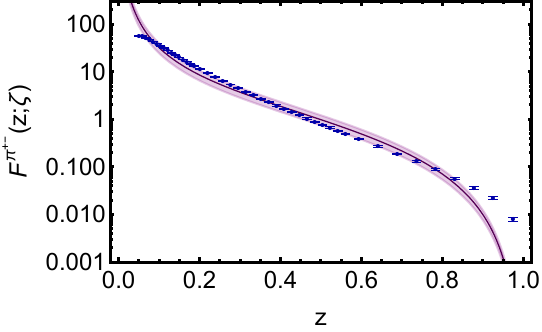} &
\includegraphics[clip, width=0.4\columnwidth]{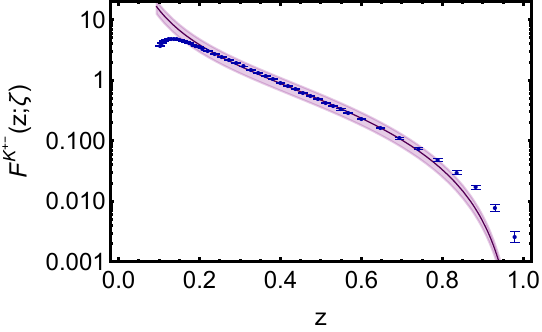} 
\end{tabular}
\caption{\label{zdepmult}
Hadron multiplicity structure function in Eq.\,\eqref{FhNorm} at $\zeta = 10.5\,$GeV -- solid purple curves with associated uncertainty bands -- compared with data drawn from Ref.\,\cite[BaBar]{BaBar:2013yrg}.
{\sf Panel A}.  $h = \pi^\pm$.
{\sf Panel B}.  $h = K^\pm$.
%
%
}
\end{figure*}

\section{Conclusions}

Exploiting the CSMs and DLY relation, we gave a unified prediction for the pion and kaon FFs of all parton species. An important feature of this approach is that the predictions are consistent with all QCD-based expectations for behavior on the endpoint domains $z\simeq 0, 1$, \emph{e.g}., nonsinglet FFs vanish at $z=0$ and singlet FFs diverge faster than $1/z$. By contrast, phenomenological fits of FFs from experimental data~\cite{Hirai:2007cx, deFlorian:2014xna, Bertone:2017tyb, Soleymaninia:2020bsq, Moffat:2021dji, AbdulKhalek:2022laj, Gao:2024dbv} are
mutually inconsistent on a large domain of $z$, particularly on $z \lesssim 0.5$. In addition, they fail to conform with expected endpoint behavior, \emph{e.g}., with singlet FFs that satisfy $z D_{\rm singlet}(z,\zeta_2)<\infty$ on $z\simeq 0$, whereupon glue and sea contributions should lead to divergences.

Turning to physical observables, our framework provides concrete predictions for hadron multiplicities in $e^+e^-$ annihilation. The predictions reveal SU(3) flavor symmetry breaking in the charged-to-neutral kaon multiplicity ratio, which is significant at lower reaction energies, $\zeta \approx 3 m_p$, and decreases as the energy increases. Furthermore, our predictions for the $z$ dependent $\pi^\pm$ and $K^\pm$ multiplicities exhibit good quantitative agreement with available experimental data on the domain $z \in [0.1, 0.7]$. These results validate the utility of this framework.  Extensions to proton and heavy-quark fragmentation currently underway.

\section*{Acknowledgments}
This contribution is based on Refs.\,\cite{Xing:2023pms, Xing:2025eip} below and I would like to thank my collaborators on these projects.
Work supported by
National Natural Science Foundation of China (grant no.\ 12135007).


\begin{thebibliography}{99}

\bibitem[{Field and Feynman(1977)}]{Field:1976ve}
\bibinfo{author}{R.~D. Field}, \bibinfo{author}{R.~P. Feynman},
  \bibinfo{title}{{Quark Elastic Scattering as a Source of High Transverse
  Momentum Mesons}}, \bibinfo{journal}{Phys. Rev. D} \bibinfo{volume}{15}
  (\bibinfo{year}{1977}) \bibinfo{pages}{2590--2616}.

\bibitem[{Field and Feynman(1978)}]{Field:1977fa}
\bibinfo{author}{R.~D. Field}, \bibinfo{author}{R.~P. Feynman},
  \bibinfo{title}{{A Parametrization of the Properties of Quark Jets}},
  \bibinfo{journal}{Nucl. Phys. B} \bibinfo{volume}{136} (\bibinfo{year}{1978})
  \bibinfo{pages}{1--76}.

\bibitem[{Altarelli(1982)}]{Altarelli:1981ax}
\bibinfo{author}{G.~Altarelli}, \bibinfo{title}{{Partons in Quantum
  Chromodynamics}}, \bibinfo{journal}{Phys. Rept.} \bibinfo{volume}{81}
  (\bibinfo{year}{1982}) \bibinfo{pages}{1--129}.

\bibitem[{Ellis et~al.(1991)Ellis, Stirling, and Webber}]{Ellis:1991qj}
\bibinfo{author}{R.~K. Ellis}, \bibinfo{author}{W.~J. Stirling},
  \bibinfo{author}{B.~R. Webber}, \bibinfo{title}{{\mbox{$\;$}QCD and collider
  physics}}, \bibinfo{publisher}{Cambridge University Press, Cambridge, UK},
  \bibinfo{year}{1991}.

\bibitem[{Metz and Vossen(2016)}]{Metz:2016swz}
\bibinfo{author}{A.~Metz}, \bibinfo{author}{A.~Vossen}, \bibinfo{title}{{Parton
  Fragmentation Functions}}, \bibinfo{journal}{Prog. Part. Nucl. Phys.}
  \bibinfo{volume}{91} (\bibinfo{year}{2016}) \bibinfo{pages}{136--202}.

\bibitem[{Chen et~al.(2023)Chen, Liu, Song, and Wei}]{Chen:2023kqw}
\bibinfo{author}{K.-B. Chen}, \bibinfo{author}{T.~Liu}, \bibinfo{author}{Y.-K.
  Song}, \bibinfo{author}{S.-Y. Wei}, \bibinfo{title}{{Several Topics on
  Transverse Momentum-Dependent Fragmentation Functions}},
  \bibinfo{journal}{Particles} \bibinfo{volume}{6}~(\bibinfo{number}{2})
  (\bibinfo{year}{2023}) \bibinfo{pages}{515--545}.

\bibitem[{Aguilar et~al.(2019)}]{Aguilar:2019teb}
\bibinfo{author}{A.~C. Aguilar}, et~al., \bibinfo{title}{{Pion and Kaon
  Structure at the Electron-Ion Collider}}, \bibinfo{journal}{Eur. Phys. J. A}
  \bibinfo{volume}{55} (\bibinfo{year}{2019}) \bibinfo{pages}{190}.

\bibitem[{Ablikim et~al.(2020)}]{BESIII:2020nme}
\bibinfo{author}{M.~Ablikim}, et~al., \bibinfo{title}{{Future Physics Programme
  of BESIII}}, \bibinfo{journal}{Chin. Phys. C}
  \bibinfo{volume}{44}~(\bibinfo{number}{4}) (\bibinfo{year}{2020})
  \bibinfo{pages}{040001}.

\bibitem[{Anderle et~al.(2021)}]{Anderle:2021wcy}
\bibinfo{author}{D.~P. Anderle}, et~al., \bibinfo{title}{{Electron-ion collider
  in China}}, \bibinfo{journal}{Front. Phys. (Beijing)}
  \bibinfo{volume}{16}~(\bibinfo{number}{6}) (\bibinfo{year}{2021})
  \bibinfo{pages}{64701}.

\bibitem[{Arrington et~al.(2021)}]{Arrington:2021biu}
\bibinfo{author}{J.~Arrington}, et~al., \bibinfo{title}{{Revealing the
  structure of light pseudoscalar mesons at the electron\textendash{}ion
  collider}}, \bibinfo{journal}{J. Phys. G} \bibinfo{volume}{48}
  (\bibinfo{year}{2021}) \bibinfo{pages}{075106}.

\bibitem[{Schnell(2022)}]{Schnell:2022nlf}
\bibinfo{author}{G.~Schnell}, \bibinfo{title}{{Fragmentation Function
  Measurements from Belle}}, \bibinfo{journal}{JPS Conf. Proc.}
  \bibinfo{volume}{37} (\bibinfo{year}{2022}) \bibinfo{pages}{020110}.

\bibitem[{Quintans(2022)}]{Quintans:2022utc}
\bibinfo{author}{C.~Quintans}, \bibinfo{title}{{The New AMBER Experiment at the
  CERN SPS}}, \bibinfo{journal}{Few Body Syst.}
  \bibinfo{volume}{63}~(\bibinfo{number}{4}) (\bibinfo{year}{2022})
  \bibinfo{pages}{72}.

\bibitem[{Hirai et~al.(2007)Hirai, Kumano, Nagai, and Sudoh}]{Hirai:2007cx}
\bibinfo{author}{M.~Hirai}, \bibinfo{author}{S.~Kumano}, \bibinfo{author}{T.~H.
  Nagai}, \bibinfo{author}{K.~Sudoh}, \bibinfo{title}{{Determination of
  fragmentation functions and their uncertainties}}, \bibinfo{journal}{Phys.
  Rev. D} \bibinfo{volume}{75} (\bibinfo{year}{2007}) \bibinfo{pages}{094009}.

\bibitem[{de~Florian et~al.(2015)de~Florian, Sassot, Epele, Hern\'andez-Pinto,
  and Stratmann}]{deFlorian:2014xna}
\bibinfo{author}{D.~de~Florian}, \bibinfo{author}{R.~Sassot},
  \bibinfo{author}{M.~Epele}, \bibinfo{author}{R.~J. Hern\'andez-Pinto},
  \bibinfo{author}{M.~Stratmann}, \bibinfo{title}{{Parton-to-Pion Fragmentation
  Reloaded}}, \bibinfo{journal}{Phys. Rev. D}
  \bibinfo{volume}{91}~(\bibinfo{number}{1}) (\bibinfo{year}{2015})
  \bibinfo{pages}{014035}.

\bibitem[{Bertone et~al.(2017)Bertone, Carrazza, Hartland, Nocera, and
  Rojo}]{Bertone:2017tyb}
\bibinfo{author}{V.~Bertone}, \bibinfo{author}{S.~Carrazza},
  \bibinfo{author}{N.~P. Hartland}, \bibinfo{author}{E.~R. Nocera},
  \bibinfo{author}{J.~Rojo}, \bibinfo{title}{{A determination of the
  fragmentation functions of pions, kaons, and protons with faithful
  uncertainties}}, \bibinfo{journal}{Eur. Phys. J. C}
  \bibinfo{volume}{77}~(\bibinfo{number}{8}) (\bibinfo{year}{2017})
  \bibinfo{pages}{516}.

\bibitem[{Soleymaninia et~al.(2021)Soleymaninia, Goharipour, Khanpour, and
  Spiesberger}]{Soleymaninia:2020bsq}
\bibinfo{author}{M.~Soleymaninia}, \bibinfo{author}{M.~Goharipour},
  \bibinfo{author}{H.~Khanpour}, \bibinfo{author}{H.~Spiesberger},
  \bibinfo{title}{{Simultaneous extraction of fragmentation functions of light
  charged hadrons with mass corrections}}, \bibinfo{journal}{Phys. Rev. D}
  \bibinfo{volume}{103}~(\bibinfo{number}{5}) (\bibinfo{year}{2021})
  \bibinfo{pages}{054045}.

\bibitem[{Moffat et~al.(2021)Moffat, Melnitchouk, Rogers, and
  Sato}]{Moffat:2021dji}
\bibinfo{author}{E.~Moffat}, \bibinfo{author}{W.~Melnitchouk},
  \bibinfo{author}{T.~C. Rogers}, \bibinfo{author}{N.~Sato},
  \bibinfo{title}{{Simultaneous Monte~Carlo analysis of parton densities and
  fragmentation functions}}, \bibinfo{journal}{Phys. Rev. D}
  \bibinfo{volume}{104}~(\bibinfo{number}{1}) (\bibinfo{year}{2021})
  \bibinfo{pages}{016015}.

\bibitem[{Abdul~Khalek et~al.(2022)Abdul~Khalek, Bertone, Khoudli, and
  Nocera}]{AbdulKhalek:2022laj}
\bibinfo{author}{R.~Abdul~Khalek}, \bibinfo{author}{V.~Bertone},
  \bibinfo{author}{A.~Khoudli}, \bibinfo{author}{E.~R. Nocera},
  \bibinfo{title}{{Pion and kaon fragmentation functions at
  next-to-next-to-leading order}}, \bibinfo{journal}{Phys. Lett. B}
  \bibinfo{volume}{834} (\bibinfo{year}{2022}) \bibinfo{pages}{137456}.

\bibitem[{Gao et~al.(2024)Gao, Liu, Shen, Xing, and Zhao}]{Gao:2024dbv}
\bibinfo{author}{J.~Gao}, \bibinfo{author}{C.~Liu}, \bibinfo{author}{X.~Shen},
  \bibinfo{author}{H.~Xing}, \bibinfo{author}{Y.~Zhao}, \bibinfo{title}{{Global
  analysis of fragmentation functions to charged hadrons with high-precision
  data from the LHC}}, \bibinfo{journal}{Phys. Rev. D}
  \bibinfo{volume}{110}~(\bibinfo{number}{11}) (\bibinfo{year}{2024})
  \bibinfo{pages}{114019}.
  
\bibitem[{Roberts et~al.(2021)Roberts, Richards, Horn, and
  Chang}]{Roberts:2021nhw}
\bibinfo{author}{C.~D. Roberts}, \bibinfo{author}{D.~G. Richards},
  \bibinfo{author}{T.~Horn}, \bibinfo{author}{L.~Chang},
  \bibinfo{title}{{Insights into the emergence of mass from studies of pion and
  kaon structure}}, \bibinfo{journal}{Prog. Part. Nucl. Phys.}
  \bibinfo{volume}{120} (\bibinfo{year}{2021}) \bibinfo{pages}{103883}.

\bibitem[{Binosi(2022)}]{Binosi:2022djx}
\bibinfo{author}{D.~Binosi}, \bibinfo{title}{{Emergent Hadron Mass in Strong
  Dynamics}}, \bibinfo{journal}{Few Body Syst.}
  \bibinfo{volume}{63}~(\bibinfo{number}{2}) (\bibinfo{year}{2022})
  \bibinfo{pages}{42}.

\bibitem[{Ding et~al.(2023)Ding, Roberts, and Schmidt}]{Ding:2022ows}
\bibinfo{author}{M.~Ding}, \bibinfo{author}{C.~D. Roberts},
  \bibinfo{author}{S.~M. Schmidt}, \bibinfo{title}{{Emergence of Hadron Mass
  and Structure}}, \bibinfo{journal}{Particles}
  \bibinfo{volume}{6}~(\bibinfo{number}{1}) (\bibinfo{year}{2023})
  \bibinfo{pages}{57--120}.

\bibitem[{Ferreira and Papavassiliou(2023)}]{Ferreira:2023fva}
\bibinfo{author}{M.~N. Ferreira}, \bibinfo{author}{J.~Papavassiliou},
  \bibinfo{title}{{Gauge Sector Dynamics in QCD}}, \bibinfo{journal}{Particles}
  \bibinfo{volume}{6}~(\bibinfo{number}{1}) (\bibinfo{year}{2023})
  \bibinfo{pages}{312--363}.

\bibitem[{Raya et~al.(2024)Raya, Bashir, Binosi, Roberts, and
  Rodr\'\i{}guez-Quintero}]{Raya:2024ejx}
\bibinfo{author}{K.~Raya}, \bibinfo{author}{A.~Bashir},
  \bibinfo{author}{D.~Binosi}, \bibinfo{author}{C.~D. Roberts},
  \bibinfo{author}{J.~Rodr\'\i{}guez-Quintero}, \bibinfo{title}{{Pseudoscalar
  Mesons and Emergent Mass}}, \bibinfo{journal}{Few Body Syst.}
  \bibinfo{volume}{65}~(\bibinfo{number}{2}) (\bibinfo{year}{2024})
  \bibinfo{pages}{60}.


\bibitem[{Xing et~al.(2024)Xing, Yao, Li, Binosi, Cui, and
  Roberts}]{Xing:2023pms}
\bibinfo{author}{H.~Y. Xing}, \bibinfo{author}{Z.~Q. Yao},
  \bibinfo{author}{B.~L. Li}, \bibinfo{author}{D.~Binosi},
  \bibinfo{author}{Z.~F. Cui}, \bibinfo{author}{C.~D. Roberts},
  \bibinfo{title}{{Developing predictions for pion fragmentation functions}},
  \bibinfo{journal}{Eur. Phys. J. C} \bibinfo{volume}{84}~(\bibinfo{number}{1})
  (\bibinfo{year}{2024}) \bibinfo{pages}{82}.


\bibitem[{Drell et~al.(1969)Drell, Levy, and Yan}]{Drell:1969jm}
\bibinfo{author}{S.~D. Drell}, \bibinfo{author}{D.~J. Levy},
  \bibinfo{author}{T.-M. Yan}, \bibinfo{title}{{A Theory of Deep Inelastic
  Lepton-Nucleon Scattering and Lepton Pair Annihilation Processes. 1.}},
  \bibinfo{journal}{Phys. Rev.} \bibinfo{volume}{187} (\bibinfo{year}{1969})
  \bibinfo{pages}{2159--2171}.

\bibitem[{Drell et~al.(1970{\natexlab{a}})Drell, Levy, and Yan}]{Drell:1969wd}
\bibinfo{author}{S.~D. Drell}, \bibinfo{author}{D.~J. Levy},
  \bibinfo{author}{T.-M. Yan}, \bibinfo{title}{{A Theory of Deep Inelastic
  Lepton Nucleon Scattering and Lepton Pair Annihilation Processes. 2. Deep
  Inelastic electron Scattering}}, \bibinfo{journal}{Phys. Rev. D}
  \bibinfo{volume}{1} (\bibinfo{year}{1970}{\natexlab{a}})
  \bibinfo{pages}{1035--1068}.

\bibitem[{Drell et~al.(1970{\natexlab{b}})Drell, Levy, and Yan}]{Drell:1969wb}
\bibinfo{author}{S.~D. Drell}, \bibinfo{author}{D.~J. Levy},
  \bibinfo{author}{T.-M. Yan}, \bibinfo{title}{{A Theory of Deep Inelastic
  Lepton-Nucleon Scattering and Lepton Pair Annihilation Processes. 3. Deep
  Inelastic electron-Positron Annihilation}}, \bibinfo{journal}{Phys. Rev. D}
  \bibinfo{volume}{1} (\bibinfo{year}{1970}{\natexlab{b}})
  \bibinfo{pages}{1617--1639}.

\bibitem[{Gribov and Lipatov(1972{\natexlab{a}})}]{Gribov:1972ri}
\bibinfo{author}{V.~Gribov}, \bibinfo{author}{L.~Lipatov},
  \bibinfo{title}{{Deep inelastic e p scattering in perturbation theory}},
  \bibinfo{journal}{Sov. J. Nucl. Phys.} \bibinfo{volume}{15}
  (\bibinfo{year}{1972}{\natexlab{a}}) \bibinfo{pages}{438--450}.

\bibitem[{Gribov and Lipatov(1972{\natexlab{b}})}]{Gribov:1972rt}
\bibinfo{author}{V.~N. Gribov}, \bibinfo{author}{L.~N. Lipatov},
  \bibinfo{title}{{e+ e- pair annihilation and deep inelastic e p scattering in
  perturbation theory}}, \bibinfo{journal}{Sov. J. Nucl. Phys.}
  \bibinfo{volume}{15} (\bibinfo{year}{1972}{\natexlab{b}})
  \bibinfo{pages}{675--684}.

\bibitem[{Salm\`e(2022)}]{Salme:2022eoy}
\bibinfo{author}{G.~Salm\`e}, \bibinfo{title}{{Explaining mass and spin in the
  visible matter: the next challenge}}, \bibinfo{journal}{J. Phys. Conf. Ser.}
  \bibinfo{volume}{2340}~(\bibinfo{number}{1}) (\bibinfo{year}{2022})
  \bibinfo{pages}{012011}.

\bibitem[{Krein(2023)}]{Krein:2023azg}
\bibinfo{author}{G.~Krein}, \bibinfo{title}{{Femtoscopy of the Matter
  Distribution in the Proton}}, \bibinfo{journal}{Few Body Syst.}
  \bibinfo{volume}{64}~(\bibinfo{number}{3}) (\bibinfo{year}{2023})
  \bibinfo{pages}{42}.


\bibitem[{Cui et~al.(2020{\natexlab{a}})Cui, Ding, Gao, Raya, Binosi, Chang,
  Roberts, Rodr\'{\i}guez-Quintero, and Schmidt}]{Cui:2020tdf}
\bibinfo{author}{Z.-F. Cui}, \bibinfo{author}{M.~Ding},
  \bibinfo{author}{F.~Gao}, \bibinfo{author}{K.~Raya},
  \bibinfo{author}{D.~Binosi}, \bibinfo{author}{L.~Chang},
  \bibinfo{author}{C.~D. Roberts},
  \bibinfo{author}{J.~Rodr\'{\i}guez-Quintero}, \bibinfo{author}{S.~M.
  Schmidt}, \bibinfo{title}{{Kaon and pion parton distributions}},
  \bibinfo{journal}{Eur. Phys. J. C} \bibinfo{volume}{80}
  (\bibinfo{year}{2020}{\natexlab{a}}) \bibinfo{pages}{1064}.



\bibitem[]{Xing:2025eip}
\bibinfo{author}{H.~Y.~Xing}, \bibinfo{author}{W.~H.~Bian},
  \bibinfo{author}{Z.~F.~Cui}, \bibinfo{author}{C.~D.~Roberts},
  \bibinfo{title}{{Kaon and pion fragmentation functions}},
  \bibinfo{journal}{Eur. Phys. J. C} \bibinfo{volume}{85}~(\bibinfo{number}{11})
  (\bibinfo{year}{2025}) \bibinfo{pages}{1305}.









\bibitem[{Ablikim et~al.(2025)}]{BESIII:2025mbc}
\bibinfo{author}{M.~Ablikim}, et~al., \bibinfo{title}{{Single Inclusive
  $\pi^\pm$ and $K^\pm$ Production in $e^+e^-$ Annihilation at center-of-mass
  Energies from 2.000 to 3.671GeV}}, 
  \bibinfo{journal}{Phys. Rev. Lett.}
  \bibinfo{volume}{135} (\bibinfo{year}{2025}) \bibinfo{pages}{151901}.

\bibitem[{Aihara et~al.(1984{\natexlab{a}})}]{TPCTwoGamma:1983lrv}
\bibinfo{author}{H.~Aihara}, et~al., \bibinfo{title}{{Charged hadron production
  in $e^+e^-$ annihilation at 29-GeV}}, \bibinfo{journal}{Phys. Rev. Lett.}
  \bibinfo{volume}{52} (\bibinfo{year}{1984}{\natexlab{a}})
  \bibinfo{pages}{577}.

\bibitem[{Aihara et~al.(1984{\natexlab{b}})}]{TPCTwoGamma:1984eoj}
\bibinfo{author}{H.~Aihara}, et~al., \bibinfo{title}{{$K^{*0}$ and $K^0_s$
  meson production in $e^+e^-$ annihilations at 29-GeV}},
  \bibinfo{journal}{Phys. Rev. Lett.} \bibinfo{volume}{53}
  (\bibinfo{year}{1984}{\natexlab{b}}) \bibinfo{pages}{2378}.

\bibitem[{Braunschweig et~al.(1989)}]{TASSO:1988jma}
\bibinfo{author}{W.~Braunschweig}, et~al., \bibinfo{title}{{Pion, Kaon and
  Proton Cross-sections in $e^+ e^-$ Annihilation at 34-{GeV} and 44-{GeV}
  Center-of-mass Energy}}, \bibinfo{journal}{Z. Phys. C} \bibinfo{volume}{42}
  (\bibinfo{year}{1989}) \bibinfo{pages}{189}.

\bibitem[{Abreu et~al.(2000)}]{DELPHI:2000ahn}
\bibinfo{author}{P.~Abreu}, et~al., \bibinfo{title}{{Charged and identified
  particles in the hadronic decay of W bosons and in e+ e- ---\ensuremath{>} q
  anti-q from 130-GeV to 200-GeV}}, \bibinfo{journal}{Eur. Phys. J. C}
  \bibinfo{volume}{18} (\bibinfo{year}{2000}) \bibinfo{pages}{203--228},
  \bibinfo{note}{[Erratum: Eur. Phys. J. C 25, 493 (2002)]}.

\bibitem[{Lees et~al.(2013)}]{BaBar:2013yrg}
\bibinfo{author}{J.~P. Lees}, et~al., \bibinfo{title}{{Production of charged
  pions, kaons, and protons in $e^+e^-$ annihilations into hadrons at
  $\sqrt{s}$=10.54 GeV}}, \bibinfo{journal}{Phys. Rev. D} \bibinfo{volume}{88}
  (\bibinfo{year}{2013}) \bibinfo{pages}{032011}.

\end{thebibliography}
\end{document}